\documentclass{aa}  
\usepackage{graphicx}
\usepackage{xcolor}
\usepackage[switch]{lineno}
\usepackage{txfonts}
\newcommand{\ltsima}{$\; \buildrel < \over \sim \;$}
\newcommand{\simlt}{\lower.5ex\hbox{\ltsima}}
\newcommand{\gtsima}{$\; \buildrel > \over \sim \;$}
\newcommand{\simgt}{\lower.5ex\hbox{\gtsima}}

\newcommand{\cgs}{${\rm erg\ cm}^{-2}\ {\rm s}^{-1}$} 
 
\newcommand{\lum}{\rm erg\ s$^{-1}$}
\newcommand{\lx}{\rm $L_{2-10~keV}$}

\def\lesssim{\mathrel{\hbox{\rlap{\hbox{\lower4pt\hbox{$\sim$}}}\hbox{$<$}}}}
\def\gtrsim{\mathrel{\hbox{\rlap{\hbox{\lower4pt\hbox{$\sim$}}}\hbox{$>$}}}}

\def\arcmin{\hbox{$^\prime$}}
\def\arcsec{\hbox{$^{\prime\prime}$}}
\def\micron{\hbox{$\mu$m}}

\def\ab1450{$AB_{1450(1+z)}$}

\def\xray{\hbox{X-ray}}

\def\oiii{\hbox{[O\ {\sc iii}}]}

\def\nh{$N_{\rm H}$}
\def\09104{IRAS~09104$+$4109}
\def\I09104{I09104}
\def\N2785{NGC~2785}

\def\edd_ratio{$\log\ L_{\rm bol}/L_{\rm Edd}$}

\def\l58{{$(\lambda L_{\lambda})_{\mbox{{\rm \scriptsize 5.8\micron}}}$}}
\def\lmir2{{$(\lambda L_{\lambda})_{\mbox{{\rm \scriptsize 12.3\micron}}}$}}
\def\s1{{S$_{\mbox{{\rm \scriptsize 3.6\micron}}}$}}
\def\irac2{{S$_{\mbox{{\rm \scriptsize 4.5\micron}}}$}}
\def\f3{{S$_{\mbox{{\rm \scriptsize 5.8\micron}}}$}}
\def\mic8{{S$_{\mbox{{\rm \scriptsize 8\micron}}}$}}
\def\f24{{F$_{\mbox{{\rm \scriptsize 24\micron}}}$}}

\def\chandra{{\it Chandra\/}}

\def\heao1{{\it HEAO-1\/}}

\def\sax{{\it Beppo}SAX\/}
\def\nustar{{\it NuSTAR\/}}

\def\xmm{{XMM-{\it Newton\/}}}
\def\suzaku{{\it Suzaku\/}}

\def\swift{{\it Swift\/}}
\def\integral{{\it Integral\/}}

\begin{document} 


%
%
\title{\textbf{\textit{NuSTAR}} reveals that the heavily obscured nucleus of NGC~2785 was the contaminant of IRAS~09104$+$4109 in the BeppoSAX/PDS hard X-rays}
\titlerunning{NGC~2785: a heavily Compton-thick AGN in a star-forming galaxy in the backyard}
\authorrunning{C. Vignali et al.}

\author{C. Vignali
          \inst{1,2}
          \and
          P. Severgnini
          \inst{3}
          \and
          E. Piconcelli
          \inst{4}
          \and
          G. Lanzuisi
          \inst{1,2}
          \and
          R. Gilli
          \inst{2}
          \and
          M. Mignoli
          \inst{2}
          \and
          A. Comastri
          \inst{2}
          \and
          L. Ballo
          \inst{5}
          \and
          K. Iwasawa
          \inst{6,7}
          \and
          V. La Parola
          \inst{8}
          }

   \institute{Dipartimento di Fisica e Astronomia, Alma Mater Studiorum, Universit\`a degli Studi di Bologna, Via Gobetti 93/2, I-40129 Bologna, Italy 
              \email{cristian.vignali@unibo.it}
         \and
                INAF -- Osservatorio di Astrofisica e Scienza dello Spazio di Bologna, Via Gobetti 93/3, I-40129 Bologna, Italy
         \and
                INAF -- Osservatorio Astronomico di Brera, Via Brera 28, I-20121 Milano, Italy
        \and
               INAF -- Osservatorio Astronomico di Roma, Via Frascati 33, I-00040 Monteporzio Catone, Roma, Italy
        \and
              \xmm\ Science Operations Centre, ESAC/ESA, PO Box 78, E-28691 Villanueva de la Can\~{a}da, Madrid, Spain
        \and
              Institut de Ci\`{e}ncies del Cosmos (ICCUB), Universitat de Barcelona (IEEC-UB), Mart\'{i} Franqu\`{e}s, 1, 08028 Barcelona, Spain
        \and
        ICREA, Pg. Llu\'{i}s Companys 23, 08010 Barcelona, Spain
        \and
        INAF -- Istituto di Astrofisica Spaziale e Fisica Cosmica di Palermo, Via Ugo La Malfa 153, I-90146 Palermo, Italy 
             }

 \date{Received July 9, 2018; Accepted August 6, 2018}

 
  \abstract
   {The search for heavily obscured active galactic nuclei (AGNs) has been revitalized in the last five years by \nustar, which has provided a good census and spectral characterization of a population of such objects, 
mostly at low redshift, thanks to its enhanced sensitivity above 10~keV compared to previous X-ray facilities, and its hard \xray\ imaging capabilities.}
   {We aim at demonstrating how \N2785, a local (z=0.009) star-forming galaxy, is responsible, in virtue of its heavily obscured active nucleus, for significant contamination in the non-imaging \sax/PDS data of 
the relatively nearby ($\approx17$\arcmin) quasar IRAS~09104$+$4109 (z=0.44), which was originally mis-classified as Compton thick.}
   {We analyzed $\approx71$~ks \nustar\ data of \N2785\ using the MYTorus model and provided a physical description of the \xray\ properties of the source for the first time.}
   {We found that \N2785\ hosts a heavily obscured ($N_{\rm H}\approx3\times10^{24}$~cm$^{-2}$) nucleus. 
The intrinsic \xray\ luminosity of the source, once corrected for the measured obscuration (\lx$\approx10^{42}$~\lum), is consistent within a factor of a few with predictions based on the source mid-infrared flux using widely adopted correlations from the literature.}
   {Based on \nustar\ data and previous indications from the Neil Gehrels \swift\ Observatory (BAT instrument), we confirm that \N2785, because of its hard \xray\ emission and spectral shape, was responsible for at least one third of the 20--100~keV emission  observed using the PDS instrument onboard \sax, originally completely associated with IRAS~09104$+$4109. Such emission led to the erroneous classification of this source as a Compton-thick quasar, while it is now recognized as Compton thin.}


   \maketitle
%
\section{Introduction}
\label{introduction}
Currently, there is a wide consensus on the fact that a fraction of $\sim$50\% of Seyfert~2 galaxies in the local 
Universe are obscured in the \xray\ band by column densities of the order of, or larger than, the inverse of the Thomson cross-section 
($N_H\ge \sigma_T^{-1} \simeq 1.5 \times 10^{24}$~cm$^{-2}$); these sources are called Compton thick (CT; e.g., \citealt{matt2000, guainazzi2005}).  
If the optical depth ($\tau = N_H\sigma_T$) for Compton scattering does not exceed values of the order of a few, \xray\ photons with energies 
higher than 10--15~keV are able to penetrate the obscuring material and reach the observer. For higher values of $\tau$, the whole \xray\ spectrum is 
depressed by Compton down-scattering, and the \xray\ photons are effectively trapped by the obscuring material irrespective of their energy. 
The former class of sources (mildly CT) can be efficiently detected by \xray\ instruments sensitive above 10~keV, while the nature of the latter (heavily CT) 
can be inferred through indirect arguments, such as the presence of a strong iron K$\alpha$ line over a flat reflected continuum, or via observations/selections 
at other wavelengths (e.g., in the mid infrared; see e.g., \citealt{vignali2014}, and references therein). 

The search for and the study of the physical properties of CT active galactic nuclei (AGNs) is relevant to understand the evolution of accreting super-massive black holes (SMBHs). 
In particular, mildly CT AGNs are the most promising candidates to explain the residual (i.e., not yet resolved) spectrum of the cosmic \xray\ background 
around its 30~keV peak (e.g., \citealt{gilli2007, treister2009, ballantyne2011, akylas2012}). 
The existence of CT objects indicates that we may be missing a non-negligible fraction of the accretion power in the Universe and of the baryonic 
matter locked in SMBHs \citep{marconi2004, comastri2015}. Because of the highly depressed emission (due to photoelectric absorption) at energies below $\approx$10~keV, 
the detection of CT AGN is often still a challenge.  

The hard X-ray band up to $\approx$100--150~keV is potentially appropriate to provide an almost unbiased census of obscured AGNs, since both Compton-thin 
and CT AGN (with column densities up to $\approx10^{25}$~cm$^{-2}$) can be detected by hard \xray\ instruments; however, the sensitivity of the 
Neil Gehrels \swift\ Observatory/BAT and \integral/IBIS surveys ($\approx10^{-11}$~\cgs; e.g., \citealt{tueller2008, beckmann2009, burlon2011, malizia2012}) 
limits this kind of investigation to the local Universe and results in a few per cent of the XRB being resolved into discrete sources at its 
peak around 20--30~keV (e.g., \citealt{ajello2012, vasudevan2013}). These studies often require follow-up observations with more sensitive 
instruments to define the source properties in a more exhaustive and comprehensive way (e.g., \citealt{severgnini2011}). 

Over the last five years \nustar, thanks to its much higher sensitivity (a factor of $\approx$100 above 10~keV) compared to the hard \xray\ instruments cited above, coupled with  
focusing optics in the hard band, has extended the search for heavily obscured AGNs up to z$\sim$3 for a limited number of sources 
\citep{delmoro2017, lansbury2017a} and is providing a more complete census than before of the obscured AGN population at low redshift 
(e.g., \citealt{balocovic2014, gandhi2014, lansbury2015, ricci2015, masini2016, boorman2016, lansbury2017b, marchesi2018, zappacosta2018}). 

Here we present the \xray\ properties of the heavily obscured nucleus of \N2785, a disky galaxy at $z$=0.009, derived from a 71~ks \nustar\ observation. 
Our interest in this source takes its origin primarily from \citet[][V11]{vignali2011}. In that work, the analysis of the 54-month \swift/BAT map in the 15--30~keV band 
suggested that the hard \xray\ emission, detected by the non-imaging PDS instrument onboard \sax\ during the observation of the luminous quasar IRAS~09104$+$4109 (hereafter I09104, at $z$=0.442) and 
originally associated with I09104 \citep{franceschini2000}, was most likely spatially coincident with \N2785. The distance between I09104 and \N2785\ positions is about 17\arcmin\ (see Fig.~2 of V11). The intensity of the hard \xray\ emission tentatively detected ($\approx3\sigma$) by \sax/PDS and associated with I09104, coupled with the presence of an iron K$\alpha$ emission line with large equivalent width 
(EW$\approx$1--2~keV), was originally used to classify I09104 as a CT quasar \citep{franceschini2000}, 
making {\it de facto} I09104 the prototype for  luminous, high-redshift CT quasars for many years.

Doubts on the classification of I09104 as a CT AGN were cast at first by \citet[][P07 hereafter]{piconcelli2007} using \xmm\ data, where the large iron EW measured by \sax/MECS was ascribed to the presence of blended iron 
emission lines (a neutral iron line from the AGN and highly ionized iron transitions from the hot gas of the surrounding cluster of which I09104 is the cD galaxy). 
Subsequently, \cite{chiang2013}, using \suzaku\ data, confirmed the Compton-thin nature of the absorber in I09104 (see also \citealt{farrah2016}, who collected the available data for I09104, including \nustar). 
Unfortunately,  \N2785 does not fall within the field of view of any of the available \xray\ datasets for I09104 (\chandra; see also \citealt{iwasawa2001}; \xmm, \suzaku, and \nustar), thus preventing a secure association of this source as the ``contaminant" in \sax/PDS data of I09104. 

\begin{figure}
\includegraphics[angle=0,width=0.5\textwidth]{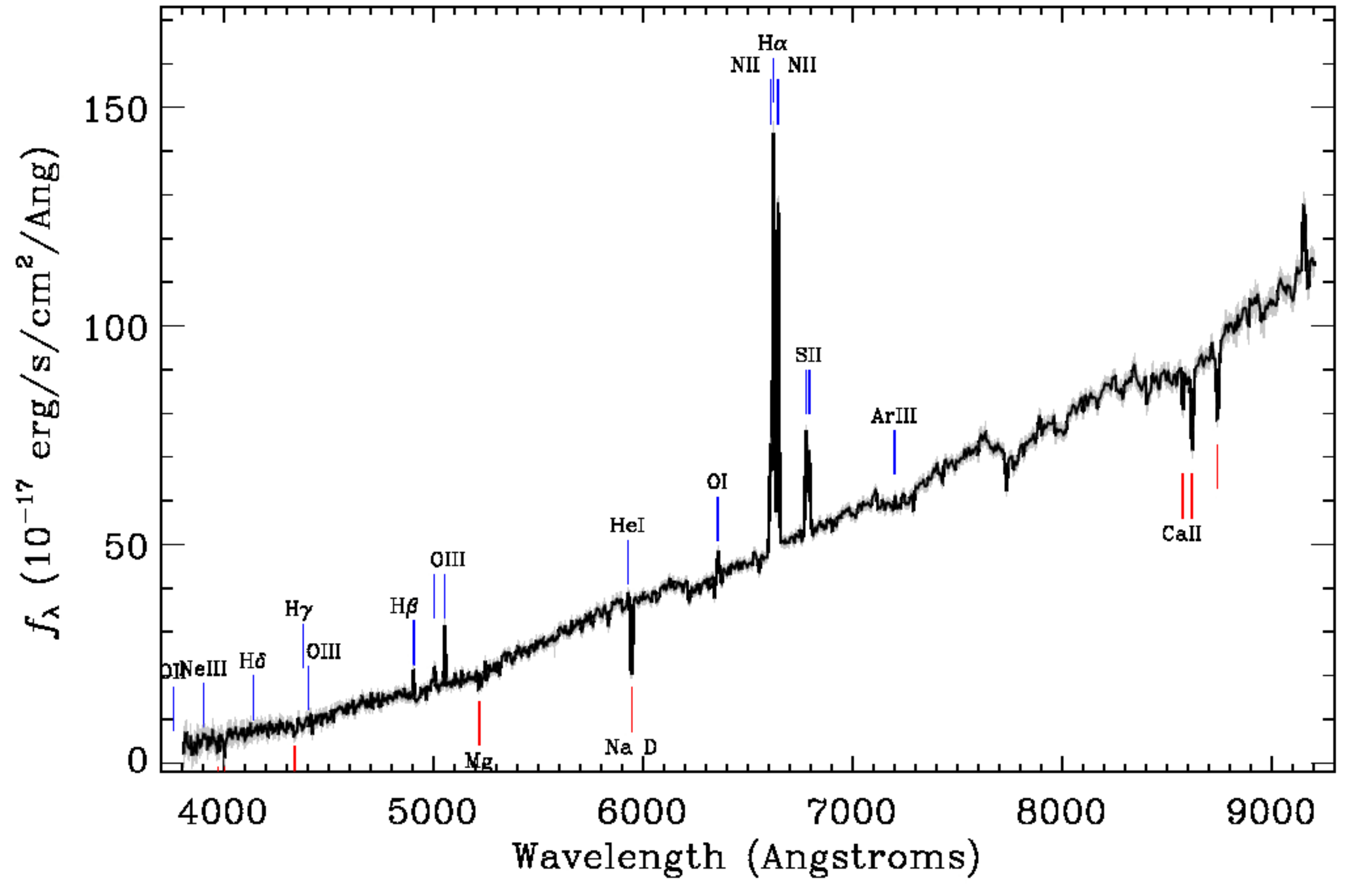}
\caption{SDSS spectrum of \N2785. The source is characterized by a very red continuum and a large H$\alpha$/H$\beta$ line ratio 
(even after correction for the host galaxy contribution), suggestive of a heavily extinguished object (see text and V11 for details). 
The main emission/absorption lines (as reported in the SDSS webpage) are labeled.}
\label{sdss_spectrum}
\end{figure}

From the optical perspective, the analysis of the Sloan Digital Sky Survey (SDSS) spectrum\footnote{http://skyserver.sdss.org/dr14/en/tools/chart/navi.aspx.} 
of \N2785\  shows a very red continuum and narrow optical emission lines 
(see Fig.~\ref{sdss_spectrum} and V11). While the presence of the AGN is inferred from the H$\alpha$/[NII] and H$\beta$/[OIII] line ratios (e.g., \citealt{veilleux1995}), the red continuum is indicative of the dominant contribution of the dust-rich host galaxy at optical wavelengths, with E(B$-$V)$\approx$1.5--1.9 assuming a Galactic extinction law \citep{seaton1979}. This represents the extinction within the host galaxy, not related to the nuclear regions. 
At face value, \N2785\ can be considered a low-redshift analog of dust-obscured galaxies at $z\approx2-3$ (e.g., \citealt{stern2014, piconcelli2015}), where a heavily 
obscured nucleus is often hidden in a star-forming galaxy. 

The present \nustar\ observation is meant to shed definitive light on the picture described above. In particular, to overcome the limited significance ($\approx$3.5$\sigma$ level) of the excess emission 
in the \swift/BAT map presented by V11, hard \xray\ imaging and sensitive data are highly required. 

In this paper we present \nustar\ data reduction, analysis, and interpretation in $\S$2, while in $\S$3 further multi-wavelength support to the nature of \N2785\ as a Compton-thick AGN is reported, providing further indication that it was a contaminant of the \sax/PDS flux observed in I09104. Conclusions are presented in $\S$4. 

The adopted cosmology consists of  $H_{0}$=70~km~s$^{-1}$~Mpc$^{-1}$, $\Omega_{\Lambda}$=0.73 and $\Omega_{M}$=0.27.
Errors are reported at the 90\%\ confidence level for one parameter of interest \citep{avni1976}, unless stated otherwise.

\section{\textbf{\textit{NuSTAR}} data}
\label{nustar_data}

\subsection{Data reduction and spectral extraction}
\label{data_reduction}
\N2785\ (RA=09:15:15.4, DEC=40:55:03; J2000) was observed by \nustar\ \citep{harrison2013} for 72~ks on May 5, 2016. The raw data were calibrated and filtered using the \nustar\ Data Analysis Software package 
({\sc NuSTARDAS}) v. 1.6.0 (included in {\sc HEASOFT} v. 6.19), the calibration files in the \nustar\ CALDB (version 20160315), and standard settings for the {\sc nupipeline} task. Inspection of the FPMA and FPMB light curves in the 3--20~keV energy range shows the presence of a short-lived ($\approx$1~ks) background flare due to solar activity, which was removed using the nustar\_filter\_lightcurve IDL script. The resulting cleaned exposure time is 71~ks for both modules. 

\begin{figure}
\centering
\includegraphics[angle=-90,width=0.48\textwidth,trim=120 10 100 50,clip]{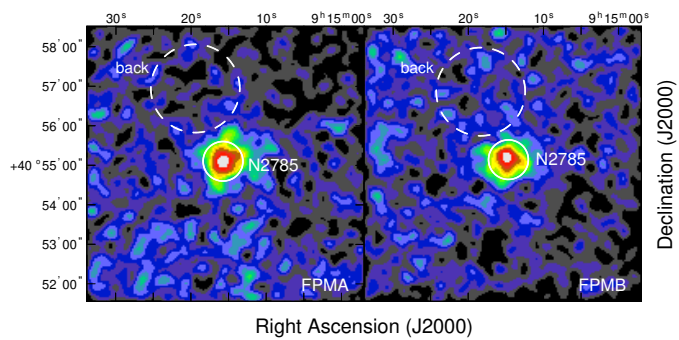}
\caption{\nustar\ images in the 4--30~keV band for the two modules (FPMA on the left, FPMB on the right). The chosen source and background extraction regions, as reported in the text, are indicated by the solid and dashed circles, respectively.}
\label{nustar_images}
\end{figure}

Source and background spectra (plus the corresponding response matrices) were extracted using the {\sc nuproducts} task. Since \N2785\ did not appear particularly bright from visual inspection of the \nustar\ images, we decided to favor the signal-to-noise ratio (S/N) in the resulting spectra by choosing a circular source extraction region (in both modules) of radius R=30\arcsec, while a larger circular (R=67\arcsec) region, free of field sources, was selected for the background (see Fig.~\ref{nustar_images}). 
The source net (i.e., background subtracted) counts in the $\approx$4--30~keV energy range are $\approx$280 and $\approx$270 in FPMA and FPMB, respectively. 
The two spectra were rebinned to have at least 20 total (i.e., source plus background) counts per bin in order to apply the $\chi^{2}$ statistics in the spectral analysis phase, performed using {\sc xspec} \citep{arnaud1996}. 
The background is not dominant in either of the two spectra, with the net source contribution being 75--80\% of the total counts over the entire \nustar\ energy band used in this work. 
However, we also checked whether the \xray\ spectral results may depend on the adopted binning method. To this goal, we binned the two spectra to a S/N of three and obtained spectral results which are fully consistent with those reported in the following analysis ($\S$\ref{spectral_results}).  
The two datasets used in the simultaneous \xray\ spectral analysis are largely consistent, as confirmed by the cross-calibration constant being close to unity. 

\subsection{X-ray spectral results}
\label{spectral_results}
A quick visual inspection of the extracted \nustar\ spectra suggests that \N2785\ is characterized by a flat \xray\ continuum coupled with a strong iron emission line, which are indicative of heavily obscured emission. To provide a preliminary, though phenomenological, description of the spectra, we adopted a model comprising an absorbed power law for the intrinsic hard \xray\ continuum, a Gaussian line for iron emission, and an additional power law to account for the emission emerging at E$<$10~keV, likely related (at least in part) to scattering (few per cent of the nuclear emission), as often observed in obscured nuclei \citep[e.g.,][]{lanzuisi2015a}.  
The fit is relatively good ($\chi^{2}$/dof= 25.3/27); the presence of extreme absorption (within the CT regime) is indicated by the derived column density of $\approx2\times10^{24}$~cm$^{-2}$ and the 
apparently strong EW ($\approx$1~keV) of the iron line. These results call for a more physically motivated modeling of the data. 

Dealing with a CT AGN candidate, to define the intrinsic continuum of \N2785\ and place constraints on the level of obscuration we used the {\sc MYTorus} model \citep{murphy_yaqoob2009}, which is based on Monte Carlo simulations; assuming a toroidal geometry for the reprocessor (uniform and cold), it self-consistently includes reflection and transmission (see e.g., \citealt{lanzuisi2015b} for further details, and Appendix~B of \citealt{lanzuisi2015a} for a comparison with other models applied to heavily obscured AGN). An additional component, parameterized by a power law, was included in the spectral fitting to account for some emission at low energy (see below for details).
We left the main parameters of the spectral fit with {\sc MYTorus} free to vary, namely the photon index, the equatorial column density of the absorber, and the inclination angle between the observer's line of sight and the symmetry axis of the torus $\theta_{\rm obs}$; the reprocessor, by construction, has a fixed half-opening angle of 60 degrees. 

\begin{figure}
\includegraphics[width=0.5\textwidth,trim=200 50 280 100,clip]{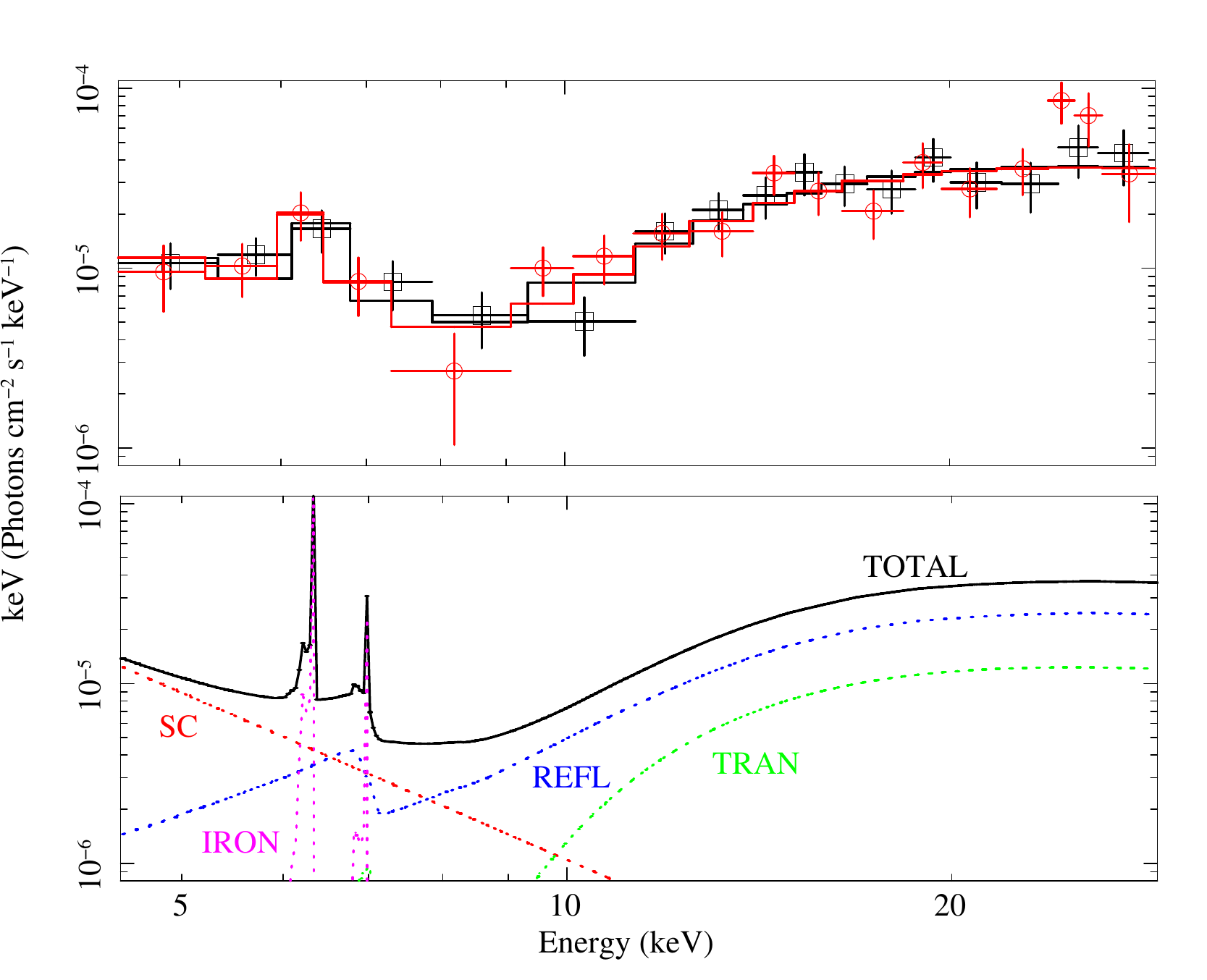}
\caption{(Top panel) \nustar\ FPMA (black) and FPMB (red) unfolded spectra of \N2785, fitted using {\sc MYTorus} and a soft component. (Bottom panel) Model components (TOTAL: best-fitting model; SC: scattering (power law) component; REFL/TRAN/IRON: reflected, transmitted and iron line components in the {\sc MYTorus} modeling (see text and \citealt{murphy_yaqoob2009}).}
\label{mytorus_spectrum}
\end{figure}

The best-fit spectral results ($\chi^{2}$/dof= 26.5/28)  shown in Fig.~\ref{mytorus_spectrum} indicate that the nucleus of \N2785\ is well parameterized by a hard power law component with $\Gamma=1.85^{+0.31}_{-0.17}$, absorbed by a column density \nh=3.0$^{+1.6}_{-0.6}\times10^{24}$~cm$^{-2}$, thus placing the source in the CT regime. 
We note that this column density value does not match -- as expected -- with the obscuration obtained converting the extinction measured in the SDSS spectrum ($\S$\ref{introduction}) using \cite{bohlin1978} and the Galactic dust-to-gas ratio. This value is of the order of 10$^{22}$~cm$^{-2}$. In fact, while the SDSS spectrum provides a measurement of the extinction on galaxy scales, the \xray-derived column density is dominated by the obscuration likely associated with the torus on much smaller (parsec) scales. 
The observer's line of sight intercepts the torus at an inclination angle of 79$^{+8}_{-10}$ degrees, suggesting an almost equatorial view of the central engine. 

The additional power law component (whose inclusion in the {\sc MYTorus} model provides a $\Delta\chi^{2}$=8.4 for two additional degrees of freedom, corresponding to an F-test probability of $\approx$98\%) turns out to be a few per cent of the nuclear one and extremely steep ($\Gamma$$\approx$4) with loose constraints, largely because of the limited photon statistics and bandpass redward of the iron line energy. Reasonably, it may be a  mixture of a scattered, AGN-related component, and thermal emission from the host galaxy, which we cannot model properly with the current data. The apparent steepness of such a component hints at the thermal emission hypothesis,  although its extrapolated 0.5--2~keV luminosity, assuming the \cite{ranalli2003} relation, would imply a star-formation rate at least one order of magnitude higher than the one derived from the spectral energy distribution (SED) fitting described in $\S$\ref{discussion_obscuration}. We also point out that a relatively good fit is still obtained if the photon index of the soft component is tied to that of the intrinsic component, with a difference in $\Delta\chi^{2}$ of $\approx$5 with respect to the best-fitting solution described above. 
However, given the quality of the \nustar\ data and the limited residuals (some deviations are present at $\approx25$~keV and are most likely instrumental), we decided not to complicate this model further. 
The observed 2--10~keV (20--30~keV) flux is $\approx2.6\times10^{-13}$~\cgs\ ($\approx5.7\times10^{-13}$~\cgs), which translates into an intrinsic (i.e., absorption-corrected) 2--10~keV luminosity of 
$\approx1.1\times10^{42}$~\lum; the ratio intrinsic/measured luminosity in the 20--30~keV band is $\approx$25. 

\section{Discussion}
\subsection{The heavily obscured nucleus of NGC~2785}
\label{discussion_obscuration}
In the following, we provide support to the heavily obscured nature of \N2785\ revealed by \nustar\ using the available multi-wavelength information. 

The analysis of its SDSS spectrum (see V11 and  $\S$\ref{introduction}) allows us to estimate the strength of the de-reddened \oiii5007\AA\ emission line which, converted into a 2--10~keV flux assuming relations by 
\citet[][or similar relations, with typical dispersions of $\approx$0.5 dex; see also \citealt{heckman2005, panessa2006, lamassa2011}]{mulchaey1994}, provides a value of $(0.9-4.7)\times10^{-11}$~\cgs\ (the range is due to the uncertainties on the E(B-V)), that is, a factor 
$\approx35-180$ higher than the one actually measured by \nustar. This is preliminary evidence, besides the \xray\ spectral analysis, that the AGN in \N2785\ is possibly strongly depressed below 10~keV. 

Furthermore, the analysis of the source SED in terms of galaxy$+$AGN emission decomposition (along the lines described in e.g., V11, and references therein) provides a first-order estimate 
of the AGN emission in the mid-IR, which can be used as a proxy of the intrinsic (e.g., 2--10~keV) AGN strength. 
The SED fitting is based on a multicomponent analysis, including stars (having the bulk of the emission in the optical/near-IR); hot dust, mainly heated by UV/optical emission due to gas accreting onto the SMBH and whose emission peaks somewhere between a few and a few tens of microns; and cold dust, principally heated by star formation, peaking around 100\micron; see \cite{fritz2006} and \cite{feltre2012} for further insights into this modeling.
Adopting mid-IR versus hard \xray\ correlations for the AGN component (see, e.g., \citealt{lanzuisi2009, gandhi2009, severgnini2012, asmus2015}, and the recent parameterizations provided by \citealt{stern2015} and \citealt{martocchia2017}), it is possible to derive the expected, intrinsic 2--10~keV luminosity. At 12\micron\  (covered by WISE data) about one quarter of \N2785\ flux is related to the AGN, while the remainder is due to the host galaxy, according to the SED decomposition described above. This translates into an intrinsic 2--10~keV luminosity of $\approx2.5\times10^{42}$~\lum\ assuming Eq.~2 of \cite{gandhi2009}.\footnote{\cite{gandhi2009} used high-resolution near-diffraction-limited observations in the mid-infrared to separate the host galaxy emission from the nuclear component in a sample of 42 local AGNs. Equation (2) refers to the subsample of 22 well-resolved sources.} 
At face value (i.e., without considering the dispersion in the used correlation, $\approx$0.2 dex), the predicted luminosity is, within a factor of approximately two, consistent with the one derived from our best-fitting \xray\ spectral solution (see  $\S$\ref{spectral_results}), further supporting the results of our analysis of \nustar\ data. 

As a further validation test, we used the AGN bolometric luminosity, obtained from the SED fitting described above, to estimate the bolometric correction in the 2--10~keV band (defined as $k_{2-10~{\rm keV}}=L_{\rm bol}/L_{2-10~{\rm keV}}$, assuming the \citealt{lusso2012} relation for spectroscopically confirmed obscured AGNs in COSMOS). For \N2785, the estimated bolometric luminosity of $\approx1.7\times10^{43}$~\lum\ implies a bolometric correction of $\approx$15. This means that the intrinsic 2--10~keV luminosity would be $\approx1.1\times10^{42}$~\lum, in agreement with the value derived from our \xray\ spectral analysis once the obscuration is taken into account. 

\subsection{The active role of NGC~2785 in contaminating BeppoSAX/PDS data}
\label{discussion_contamination}
As previously introduced, one of the goals of the current \nustar\ observation of \N2785\ was to understand whether, and at what level, its hard \xray\ emission might be responsible for the tentative detection of I09104 in \sax/PDS non-imaging data (20--100~keV flux of $1.0\pm{0.3}\times10^{-11}$~\cgs, \citealt{franceschini2000}), making it the prototype for luminous, high-redshift CT quasars for many years. In a repeat analysis of the same data carried out by P07, I09104 was reported as a 2.5$\sigma$ detection in the 15--50~keV band, with a flux of 
2.55$^{+1.90}_{-1.56}\times10^{-12}$~\cgs\ in the 20--30~keV band. 
The same authors (and, in the following years, V11; \citealt{chiang2013,farrah2016}, using multiple \xray\ datasets with I09104 as the target) have produced further evidence that I09104 was originally mis-classified as CT, and is in fact  in the Compton-thin regime (with a column density of $\approx4\times10^{23}$~cm$^{-2}$ according to the \chandra\ analysis by V11). 
However, the ``culprit" of the hard \xray\ contamination was still missing. The first piece of the puzzle towards a solution of this issue was placed by \swift/BAT, whose 54-month data in the 15--30~keV band (V11) 
suggested that \N2785\ could be the contaminant (lying at a distance of $\approx$17\arcmin\ from I09104). However, the limited S/N ($\approx3.5\sigma$ in V11, being still below the 4.8$\sigma$ detection limit in the recently published 105-month analysis; \citealt{oh2018}), combined with the poor positional accuracy in \swift/BAT data, prevented V11 from drawing firm conclusions about the association of \N2785\ with the ``excess flux" detected by \sax. 

\nustar, because of its higher sensitivity and spatial resolution than previous hard \xray\ missions, has been able to provide the ``smoking gun":  \N2785, {\it in virtue} of its heavily obscured nuclear emission emerging strongly in hard X-rays, is able to broadly justify the previously reported \sax\ flux, admittedly characterized by large uncertainties due to a marginal detection in the non-imaging PDS instrument. 
In fact, if we consider the combined 20--30~keV flux produced by \N2785\ (current \nustar\ data), and that relative to I09104 \citep[][\nustar\ data]{farrah2016}, we obtain 
a value of $(5.7+7.0)\times10^{-13}\approx1.3\times10^{-12}$~\cgs, which is consistent, within the errors, with the $\approx2.6\times10^{-12}$~\cgs\ reported by P07 from the repeat analysis of \sax/PDS data. 
If we compare the 20--100~keV band flux of I09104 in the PDS \citep{franceschini2000} with the extrapolation of I09104 and \N2785\ fluxes from \nustar\ data in the same energy range, 
$\approx6\times10^{-12}$~\cgs, we can safely say that most of the high-energy flux detected by the PDS is accounted for. Some source variability and the presence of other lower-luminosity sources in the PDS field of view (see P07 and V11) can account for the low discrepancy between the two fluxes. 
As a side note,  the angular response of the PDS collimators is $\sim$80\% at the position of \N2785\ in the \sax/PDS data of I09104 (see Fig.~2 of \citealt{frontera2007}), meaning that the expected flux contribution of \N2785\  reported above should not be significantly scaled down once the offset position of \N2785\ with respect to I09104 is taken into account. 

In the following we provide a rough estimate of the incidence of cases similar to the one described in this work "hidden" in old datasets taken by X-ray telescopes with non-focusing optics. 
In particular, we start focusing on \sax\ detections of CT AGN above 10~keV using the compilation presented in the review by \cite{comastri2004}. 
We complement this list with the one presented by \cite{dellaceca2008}. Although probably not complete, the final list is suitable to make an estimation of the fraction of sources that were originally classified as CT by \sax\ using $>$10~keV data but were probably contaminated by relatively nearby objects in the PDS (not necessarily CT as in our case). 
Five out of 25 AGNs of this list (20\%), including I09104, were not confirmed as CT using more recent datasets (in some cases comprising \nustar). 
On the one hand, \nustar\  clearly\ has a relevant role in this kind of investigation at hard X-rays compared to non-imaging instruments 
because of its focusing optics and higher sensitivity; on the other hand, good-quality \xray\ data below 10 keV are often sufficient to shed some (maybe not conclusive) light on the nature of heavily obscured AGNs even in the absence of sensitive data at higher energies. 

Recently, \cite{marchesi2018} reported that Compton-thin AGNs detected by \swift/BAT (characterized by non-focusing optics) may have been mis-classified 
as CT (40\% in the original 70-month sample) mainly because of low-quality data, leading to a general over-estimation of the true intrinsic \xray\ obscuration. 
This result, which was achieved using \nustar\ coupled with sensitive observations below 10~keV, suggests a slight revision of the fraction of CT AGNs in the \swift/BAT sample towards lower values (from 7.6\% in \citealt{ricci2015} compilation to 6\% - and, potentially, down to 4\% - in the \citealt{marchesi2018} analysis). 
In this context, given the \sax\ results discussed above, we may argue that, overall, the impact of cases such as the one described in our work (contamination in the field of view of non-imaging instruments, causing the target to be detected as CT) on the cosmic X-ray background modeling is probably limited. 

\section{Conclusions}
NGC~2785 represents a very good case of a buried AGN in a dusty galaxy, likely 
a low-redshift analog of the heavily absorbed AGN found in dust-obscured galaxies at $z\approx2-3$.  
The $\approx$70~ks \nustar\ observation has allowed, for the first time, to obtain a medium-quality \xray\ spectrum of this 
source, which was previously found to be highly extinguished on the basis of its SDSS spectrum alone. 
Based on an admittedly low S/N detection in the 54-month \swift/BAT catalog (V11), 
the source was also supposed to be the contaminant of I09104 (an obscured quasar at z=0.44) flux in the non-imaging \sax/PDS hard \xray\ data.  

Applying the {\sc MYTorus} model to the \nustar\ data has allowed us to place constraints on \N2785\ nuclear emission, which is absorbed by a 
column density of $\approx3\times10^{24}$~cm$^{-2}$, and to derive the source's intrinsic luminosity as being $\approx10^{42}$~\lum\ in the 2--10~keV band. This result places \N2785\ among the ``certified'' CT AGNs. 
Further support to the presence of heavy obscuration (i.e., to the fact that intrinsically the source is much brighter than actually observed) comes from the de-reddened \oiii\ intensity, the AGN mid-IR flux, and the source bolometric luminosity (derived from SED fitting). The claim (V11) that this source could contaminate the \sax/PDS data of I09104 is fully corroborated by the current findings. 

The results presented in this work clearly stress the need for sensitive imaging and spectroscopic observations at energies above 10~keV to 
disclose heavily obscured nuclei.  \nustar\ represents the most efficient way to provide a 
comprehensive picture of sources like \N2785\ and useful information for the adequate interpretation of their optical/mid-infrared properties.

\begin{acknowledgements}
CV thanks the referee for her/his useful comments. The authors acknowledge financial support from the Italian Space Agency (ASI) under the contracts ASI-INAF I/037/12/0 
and ASI-INAF n.2017-14-H.0. 
CV thanks M. Orlandini for a useful discussion about the response of the PDS instrument onboard \sax, and L. Zappacosta for indications about \nustar\ data analysis in the presence of flares.
\end{acknowledgements}

\end{document}